\documentstyle[epsf]{mnsty}
\def \farcs{\hbox{$.\!\!^{\prime\prime}$}}

\def \moverl{\hbox{$h_{50}~{\rm M}_\odot/{\rm L}_{{\rm B}\odot}$}}       

\title[Weak lensing study of X-ray selected clusters]
{HST large field weak lensing analysis of MS~2053-04: study of
the mass distribution and mass-to-light ratio of X-ray selected
clusters at $0.22<z<0.83$$^\star$}

\author[Hoekstra et al.]
{Henk Hoekstra$^{1,2,3}$, Marijn Franx$^4$, Konrad Kuijken$^3$, \&
	Pieter G. van Dokkum$^{5 \dagger}$\\
	$^1$~CITA, University of Toronto, 60 St. George Street,
	Toronto, M5S 3H8, Canada\\
	$^2$~Department of Astronomy, University of Toronto, 
	60 St. George Street, Toronto, M5S 3H8, Canada\\
	$^3$~Kapteyn Astronomical Institute, University of Groningen, 
        Postbus 800, 9700 AV Groningen, The Netherlands \\
	$^4$~Leiden Observatory, P.O.~Box 9513, 2300 RA Leiden, 
	The Netherlands\\
	$^5$~California Institute of Technology MS 105-24, 
	Pasadena, CA 91125, USA}

\begin{document}

\maketitle

\begin{abstract} 

We have detected the weak lensing signal induced by the cluster of
galaxies MS~2053-04 $(z=0.58)$ from a two-colour mosaic of 6 HST WFPC2
images. 

The best fit singular isothermal sphere model to the observed
tangential distortion yields an Einstein radius $r_E=6{\farcs}2\pm
1{\farcs}8$, which corresponds to a velocity dispersion of
$886^{+121}_{-139}$ km/s $(\Omega_m=0.3,~\Omega_\Lambda=0.0)$.  This
result is in good agreement with the observed velocity dispersion of
$817\pm80$ km/s from cluster members.  The observed average restframe
mass-to-light ratio within a $1~h_{50}^{-1}$ Mpc radius aperture is
$184\pm56~\moverl$. After correction for luminosity evolution to $z=0$
this value changes to $291\pm89\pm19~\moverl$ (where the first error
indicates the statistical uncertainty in the measurement of the
mass-to-light ratio, and the second error is due to the uncertainty in
luminosity evolution).

MS~2053 is the third cluster we studied using mosaics of deep WFPC2
images. For all three clusters we find good agreement between
dynamical and weak lensing velocity dispersions, in contrast to
weak lensing studies based on single WFPC2 pointings on cluster cores.
This result demonstrates the importance of wide field data.

We have compared the ensemble averaged cluster profile to the
predicted NFW profile, and find that a NFW profile can fit the
observed lensing signal well. The best fit concentration parameter is
found to be $0.79^{+0.44}_{-0.15}$ (68\% confidence) times the
predicted value from an open CDM model.

The observed mass-to-light ratios of the clusters in our sample evolve
with redshift, and are inconsistent with a constant, non-evolving,
mass-to-light ratio at the 99\% confidence level. The evolution is
consistent with the results derived from the evolution of the
fundamental plane of early type galaxies.  The resulting average
mass-to-light ratio for massive clusters at $z=0$ is found to be
$239\pm 18 \pm 9~\moverl$.

\end{abstract}

\section{Introduction}
\footnotetext[1]{Based on observations with the NASA/ESA {\it Hubble
Space Telescope} obtained at the Space Telescope Science Institute,
which is operated by the Association of Universities for Research
in Astronomy, Inc., under NASA contract NAS 5-26555} 
\footnotetext[2]{Hubble Fellow}

Observations of high redshift clusters are valuable to test our current
understanding of structure formation on cosmological scales (e.g. Eke,
Cole \& Frenk 1996; Bahcall \& Fan 1998). In particular reliable mass
estimates of these systems are important, as they provide strong
constraints on cosmological models.

The small, systematic, distortion in the shapes of background sources
induced by massive structures, known as weak gravitational lensing,
has proven to be a powerful method to measure the masses of clusters
of galaxies (for an extensive review see Mellier 1999). The weak
lensing effect allows one to reconstruct the projected surface mass
density (e.g. Kaiser \& Squires 1993) or measure the mass, without
having to rely on assumptions about the state or nature of the
deflecting matter. However, for an accurate mass estimate high number
densities of background galaxies are needed, as well as a good
estimate of their redshift distribution.

Lensing studies of high redshift clusters $(z>0.5)$ are difficult
because the lensing signal is low and most of the signal comes from
small, faint sources. These sources typically have sizes which are
comparable to the size of the PSF in ground based images. To extract
the lensing signal from such observations large corrections are
required.  For these studies HST observations have great advantage
over ground based observations because the background sources are much
better resolved, resulting in a well calibrated weak lensing signal.

In this paper we present the results of our weak lensing analysis of
the $z=0.58$ cluster of galaxies MS~2053-04. It is the third cluster
of which we studied the mass distribution based on a deep two-colour
mosaic of WFPC2 images. The other two clusters that have been studied
this way are Cl~1358+62 $(z=0.33)$ (Hoekstra et al. 1998; HFKS
hereafter), and MS~1054-03 $(z=0.83)$ (Hoekstra, Franx, \& Kuijken
2000; HFK hereafter). All three clusters have been selected on the
basis of their strong X-ray emission.

MS~2053 was detected in the Einstein Medium Sensitivity Survey (Gioia
\& Luppino 1994). It is one of the few $z>0.5$ clusters found in this
survey, and of these high redshift clusters it has the lowest X-ray
luminosity. Its X-ray luminosity\footnote[2]{Throughout this paper we
will use $H_0=50~h_{50}$km/s/Mpc, $\Omega_m=0.3$ and
$\Omega_\Lambda=0$. This gives a scale of $1''=8.2 h_{50}^{-1}$ kpc at
the distance of MS~2053.} is $L_x(2-10~{\rm keV})=(7.9\pm0.7)\times
10^{44}~h_{50}^{-2}$ ergs/s (Henry 2000). The X-ray temperature
measured by BeppoSAX is $kT=6.7^{+6.8}_{-2.3}$ keV (Della Ceca et
al. 2000). A more accurate temperature of $kT=8.1^{+3.7}_{-2.2}$ keV
has been determined from ASCA observations (Henry 2000).

Luppino \& Gioia (1992) discovered a gravitationally lensed arc in
deep images of MS~2053. The arc is located approximately 15 arcseconds
from the Brightest Cluster Galaxy (BCG). Its redshift is still
unknown.  The cluster mass distribution has been studied previously
through weak lensing by Clowe (1998) based on deep ground based
images.

We first present the results of the weak lensing analysis of MS~2053.
In section~2 we briefly discuss the data, and in section~3 the object
analysis is described.  The cluster light distribution is examined in
section~4.  In section~5 we present the weak lensing signal and the
reconstruction of the projected surface mass density. The mass and
mass-to-light ratio inferred from our analysis are presented in
section~6. In section~7 we present the combined results of a sample of
4 clusters that have analysed and calibrated in a uniform way. We
compare the weak lensing mass estimates to dynamical estimates. We
also study the average mass profile of the clusters, as well as their
mass-to-light ratios.

\begin{figure}
\begin{center}
\leavevmode
\hbox{%
\epsfxsize=\hsize
\epsffile{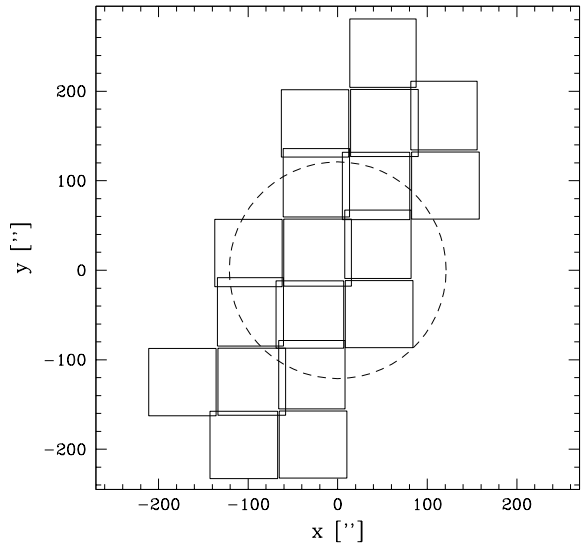}}
\begin{small}
\caption{Layout of the observed field. The six pointings are
indicated.  The dashed circle indicates an aperture with a radius of
$1~h_{50}^{-1}$ Mpc centered on the brightest cluster galaxy
(BCG). The area covered by the observations is approximately 26.5
armin$^{2}$.
\label{layout}}
\end{small}
\end{center}
\end{figure}  

\section{Data}

To study the cluster MS~2053 we use a mosaic of WFPC2 images taken
with the Hubble Space Telescope. Figure~\ref{layout} shows the layout
of the mosaic constructed from the 6 pointings of the telescope. The
cluster has been observed in two passbands. Each pointing in each
filter consists of three separate short exposures, which allows an
effective rejection of cosmic rays. The total integration time per
pointing was 3300s in the $F606W$ filter, and 3200s in the $F814W$
filter. The reduction is described in van Dokkum et al. (2001). For
the weak lensing analysis we omit the data of the Planetary Camera
because the data do not reach the same depth as the Wide Field
Camera. The total area covered by the observations is approximately
26.5 arcmin$^{2}$.

\begin{figure*}
\begin{center}
\leavevmode
\hbox{%
\epsfxsize=8cm
\epsffile{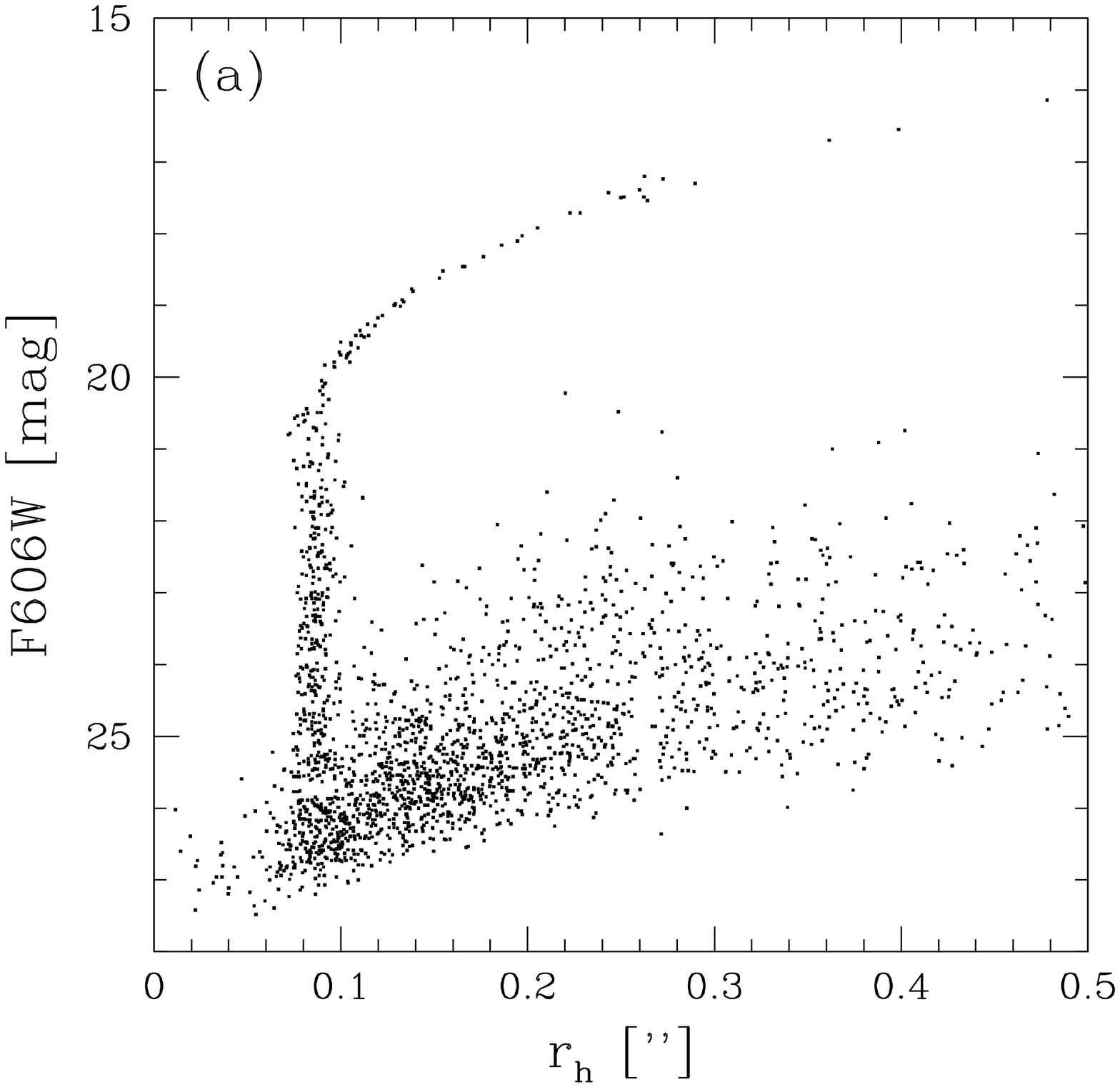}
\epsfxsize=8cm
\epsffile{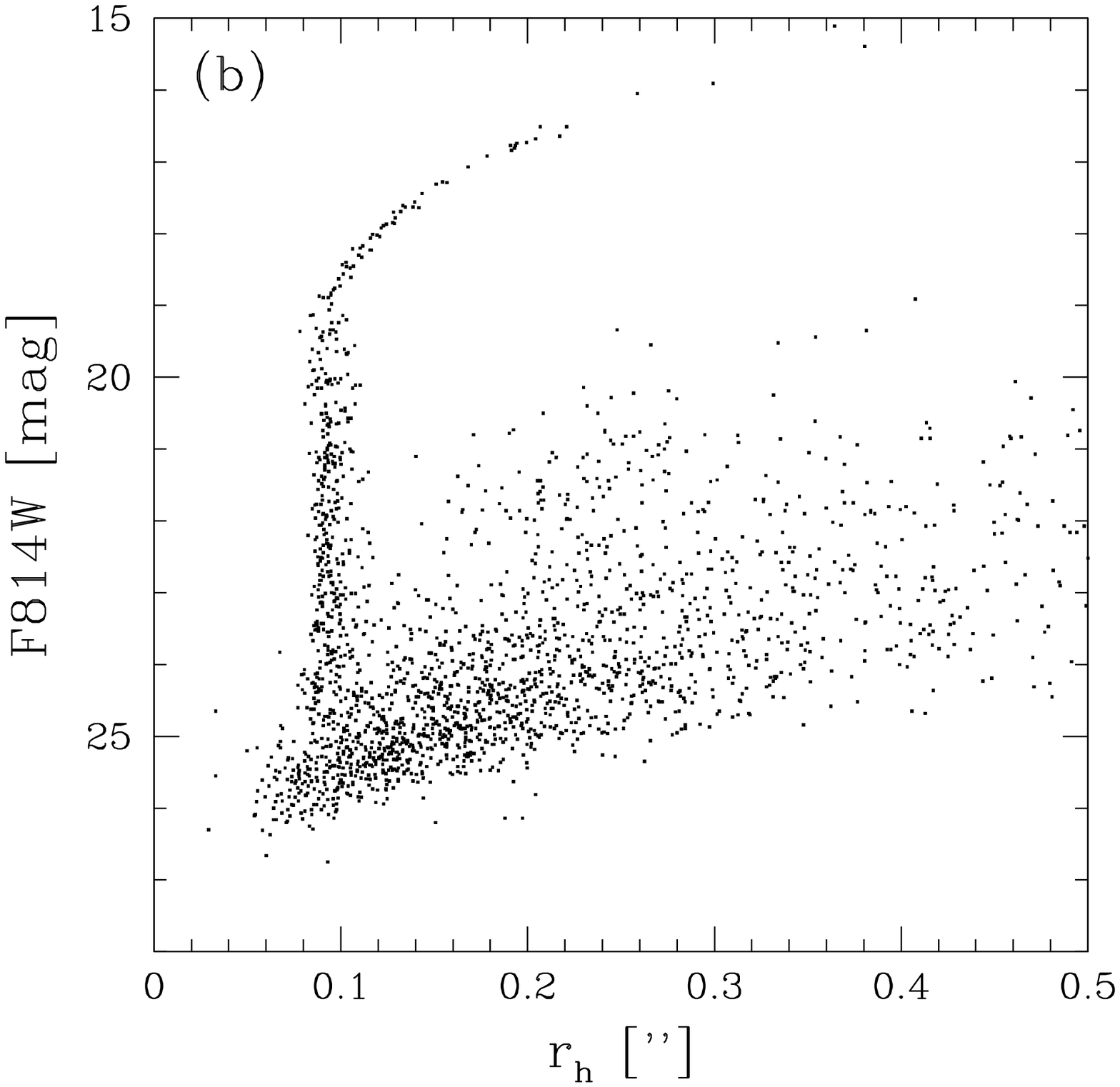}}
\begin{small}
\caption{Plot of the apparent magnitude in the $F606W$ filter (a) and
the $F814W$ filter (b) versus half light radius $r_h$. Because of the low
galactic latitude of MS~2053 many stars are found, which correspond
to the vertical sequence of points at $r_h\sim 0\farcs{1}$.
\label{sizemag}}
\end{small}
\end{center}
\end{figure*}    

\section{Object analysis}

The weak lensing analysis technique is based on that developed by
Kaiser, Squires, \& Broadhurst (1995), and Luppino \& Kaiser (1997),
with a number of modifications which are described in detail in HFKS
and HFK. We analyse each WFPC2 chip separately, and combine the object
catalogs to a master catalog once all objects have been analysed and
the appropriate corrections have been applied.

We use the hierarchical peak finding algorithm from Kaiser et
al. (1995) to find objects with a significance $>5\sigma$ over the
local sky. These are analysed, which yields estimates for their sizes,
magnitudes and shapes. As described in HFK we also estimate the error
on the shape measurements, which allows a proper weighting of the
sources. 

The resulting catalogs are inspected visually, and spurious detections,
such as diffraction spikes, HII regions in resolved galaxies, etc. are
removed.

We then identify the objects that are detected in both the $F606W$ and
$F814$ images.  For these we determine colours using the same aperture
for both filters. The aperture that is used scales with the Gaussian
scale length $r_g$ of the object. This results in a sample of 2155
objects, both galaxies and stars, with a corresponding number density
of 81 objects arcmin$^{-2}$. The objects that are detected in only one
filter are small, and faint, and as a result not useful for the weak
lensing analysis.

The magnitudes are zero-pointed to Vega, using the zero points given
the {\it HST} Data Handbook (Voit et al. 1997). Figure~\ref{sizemag}
shows the plot of the apparent magnitude versus the object half light
radius of the detected objects. 

Because of the low galactic latitude of MS~2053 many stars are found
in the observed field. These are located in the vertical sequence of
points at a half-light radius $r_h\sim 0\farcs{1}$. The brightest
stars saturate and have larger half light radii. Based on
Figure~\ref{sizemag} we select a sample of 198 moderately bright
stars.  These stars are used to study the PSF, and the results are
used to correct the shapes of the faint galaxies for PSF anisotropy
and the size of the PSF as described in HFKS.

\begin{figure}
\begin{center}
\leavevmode
\hbox{%
\epsfxsize=7.5cm
\epsffile[60 200 595 720]{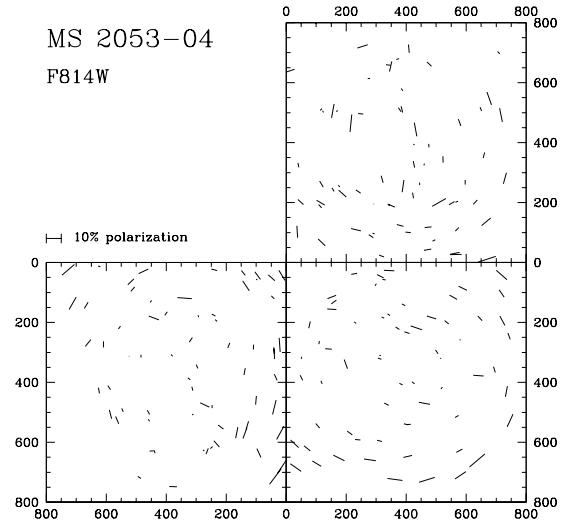}}
\begin{small}
\caption{Observed polarization of stars selected from the $F814W$ images.
The sticks indicate the direction of the major axis of the PSF, as well
as the size of the polarization. The polarizations are measured using
a Gaussian weight function with a dispersion of 0\farcs{07}. The lower
left panel corresponds to chip2, the lower right to chip 3, and the
upper right one denotes chip 4. We have omitted chip 1, which is the
planetary camera.
\label{psf}}
\end{small}
\end{center}
\end{figure}

The observed polarizations in the $F814W$ images of these stars are
presented in Figure~\ref{psf}. HFKS studied the WFPC2 PSF using observations
of the globular cluster M4, and the pattern observed here is similar to
the one presented in HFKS.

The PSF changes slightly with time, and subtraction of the M4 model
from the observations leaves systematic residuals. To improve the
model for the PSF anisotropy we fitted a modified model to
the shape parameters of the stars in the MS~2053. It is a scaled
version of the M4 model with a first order polynomial added:
$$p_i^{\rm new}=a\cdot p_i^{\rm M4} + c_0 +c_1 x + c_2 y.$$ This model
fits the observed PSF anisotropy of stars in the MS~2053
field well (the reduced $\chi^2$ of the fit is 0.98 for 179 stars).

The next step is to determine the ``pre-seeing'' shear polarizability
$P^\gamma$ (Luppino \& Kaiser 1997; HFKS). The measurements of $P^\gamma$
for individual galaxies are rather noisy, and therefore we bin the
measurements as a function of the Gaussian scale length $r_g$.

Because of the poor sampling of WFPC2 images only the shapes of galaxies
with size $r_g> 0\farcs{08}$ can be corrected reliably\footnote{The peak 
finder program provides an estimate for $r_g$ that is a factor $\sqrt{2}$
too large. This affects all objects, and therefore the limit listed
here corresponds to the 0\farcs{12} given by HFKS.}. We
select objects that have $r_g>0\farcs{08}$ and remove saturated stars
from the catalogs. After this selection the sample of galaxies
consists of 1540 galaxies analysed from the $F606W$ images, and 1545
from the $F814W$ images. We note that because of this cut not all
objects appear in both catalogs anymore.

\begin{figure}
\begin{center}
\leavevmode
\hbox{%
\epsfxsize=\hsize
\epsffile[30 180 580 680]{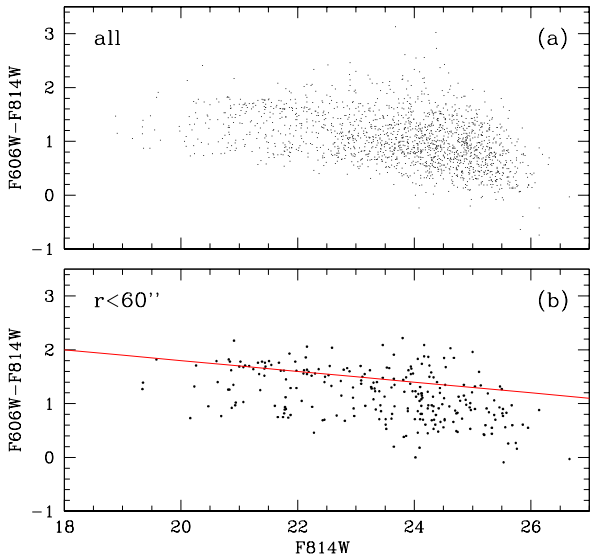}}
\begin{small}
\caption{(a) Colour-magnitude diagram of the galaxies in the full
mosaic for which colours have been determined. (b) Colour-magnitude
diagram for galaxies within 1 arcminute from the BCG. The cluster
colour-magnitude relation is better visible in this diagram, although
it is not as obvious as for other rich clusters. The line indicates the
assumed cluster colour-magnitude relation.\label{colmag}}
\end{small}
\end{center}
\end{figure}    

Finally the shapes are corrected for the camera distortion, and the
catalogs are combined into a master catalog. We use the estimated
errors on the shape measurements to combine the results from the
$F606W$ and $F814W$ images in an optimal way. The resulting catalog
includes 1677 galaxies, which corresponds to a number density of 63
galaxies arcmin$^{-2}$.

\section{Light distribution}

Figure~\ref{colmag}a shows the colour of the galaxies in the full
mosaic versus their $F814W$ magnitude. Compared to the other two
clusters for which we obtained HST mosaics, MS~2053 is less obvious
from the optical images. As a result the contrast of the cluster
colour-magnitude relation with the background is lower.  However, in
the diagram for galaxies within 1 arminute from the BCG the cluster
colour-magnitude relation can be discerned (fig.~\ref{colmag}b).

\begin{figure*}
\vspace{-0.5cm}
\begin{center}
\leavevmode
\hbox{%
\epsfxsize=0.9\hsize
\epsffile{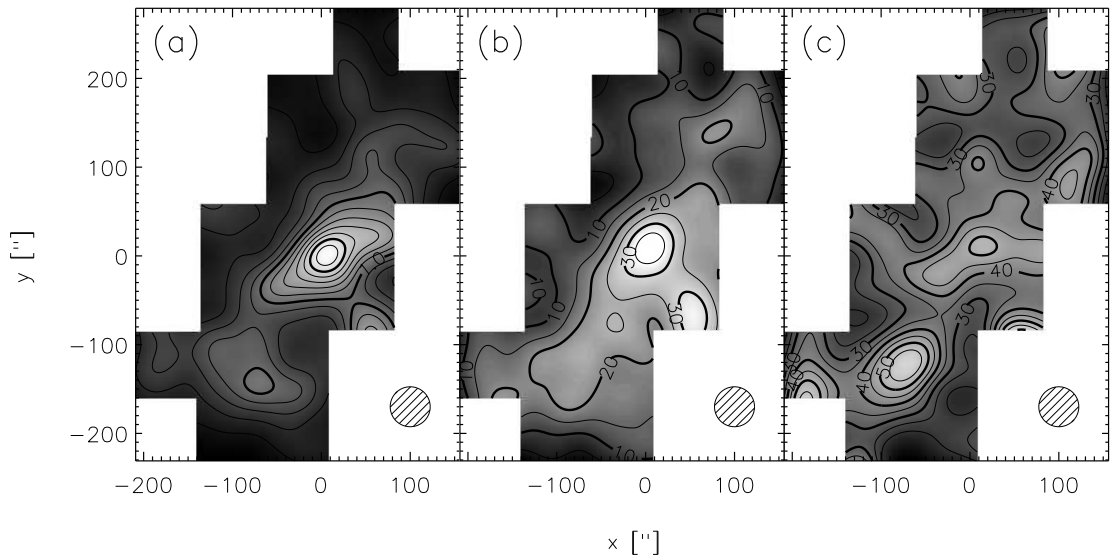}}
\vspace{-0.2cm}
\begin{small}
\caption{(a) Smoothed luminosity distribution from the sample of
cluster galaxies. The definition of the sample is described in the
text. The intervals between subsequent contours are $10^6 {\rm
L}_{B\sun}{\rm pc}^{-2}$. The position of the BCG has been used to
define the cluster centre, and corresponds to the origin of this
figure. The overdensity at the cluster centre is clearly visible.  (b)
Smoothed number density of bright galaxies ($19.5<F814W<23$). (c)
smoothed number density of faint galaxies ($23<F814W<25$). At the
position of the cluster a small overdensity is still visible. Another
overdensity of faint galaxies is visible to the lower left of the
cluster.  The number density distributions have been smoothed using a
Gaussian with a FWHM of 45'' (indicated by the shaded circle). In
figures (b) and (c) the intervals between subsequent contours are 5
galaxies arcmin$^{-2}$.
\label{lumdis}}
\end{small}
\vspace{-0.5cm}
\end{center}
\end{figure*}

To estimate the light contents of the cluster we define a sample of
cluster galaxies as follows. Down to $F814W=21$ we select
spectroscopically confirmed cluster members (Van Dokkum et al. 2001,
in preparation).  At fainter magnitudes we use the colour-magnitude
relation drawn in Figure~\ref{colmag}b. We select galaxies with
$-0.4<\Delta(F606W-F814W)<0.2$ mag relative to the cluster
colour-magnitude relation. To correct for contamination by field
galaxies we subtract the counts from the Hubble Deep Fields north and
south. The smoothed luminosity distribution of this sample is
presented in Figure~\ref{lumdis}a. In Figures~\ref{lumdis}b and c grey
scale images of the smoothed number density of bright
$(19.5<F814W<23)$ and faint $(23<F814W<25)$ galaxies are
presented. Around the position of the BCG a significant overdensity of
galaxies is detected. Figure~\ref{lumdis}a shows most clearly that the
light distribution is elongated in the direction where the arc is
found (Luppino \& Gioia 1992). Figure~\ref{lumdis}c still shows an
overdensity at the position of the cluster. Interestingly, it also
shows a clear overdensity south of the cluster. The overdensity is
caused by galaxies bluer than the cluster, but it is not clear whether
they belong to another cluster along the line of sight.

We estimate the cluster luminosity in the rest frame $B$ band. To do
so we use template spectra for a range in spectral types and compute
the corresponding pass band correction (this procedure is similar to
the method described in van Dokkum \& Franx 1996). Thus we find the
following transformation from the HST filters to the rest frame $B$
band:
$$B_z=F814W+0.47(F606W-F814W)+0.75,$$
\noindent where $B_z$ denotes the corrected $B$ band
magnitude. The luminosity is given by
$$L_B=10^{0.4(M_{B\odot}-B_z+DM+A_{F814W})} L_{B\odot},$$
\noindent where $M_{B\odot}=5.48$ is the solar absolute $B$ magnitude,
$DM$ is the distance modulus, and $A_{F814W}$ is the extinction
correction in the $F814W$ filter towards MS~2053. The redshift of
$z=0.58$ for MS~2053 gives a distance modulus of $43.14-5\log
h_{50}$. We use the dust maps from Schlegel, Finkbeiner, \& Davis
(1998) to correct for the galactic extinction. Because of the low
galactic latitude of MS~2053, we find a rather high value of
$A_{F814W}=0.15$.  We have used SExtractor (Bertin \& Arnouts 1996) to
determine total magnitudes for the galaxies.

\begin{figure}
\begin{center}
\leavevmode
\hbox{%
\epsfxsize=0.9\hsize
\epsffile{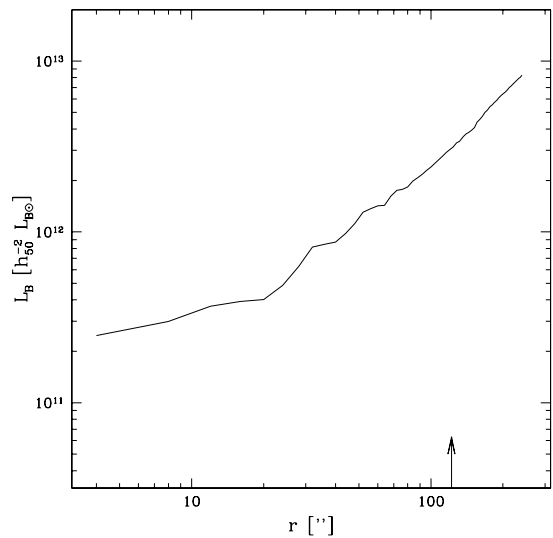}}
\begin{small}
\caption{The cumulative restframe $B$ band luminosity as a function of
radius from the cluster centre. The profile is calculated using the
sample of cluster galaxies (selected both spectroscopically and based
on colour). The arrow indicates a radius of $1~h_{50}^{-1}$ Mpc.
\label{lumprof}}
\end{small}
\end{center}
\end{figure}

The cumulative light profile as a function of distance from the
cluster centre is presented in Figure~\ref{lumprof}. The total
luminosity within an aperture of radius $1~h_{50}^{-1}$ Mpc is
$(3.1\pm 0.4)\times 10^{12}h_{50}^{-2}{\rm L}_{B\odot}$. The error in
the luminosity reflects the uncertainty in the determination of
cluster membership, and the total magnitudes measured by SExtractor.
We note, however, that the error is small compared to the uncertainty
in the weak lensing signal (section~4 and further). At large radii the
profile is rather steep. To compute the profile, we average the light
distribution in circular bins.  If the light distribution is elongated
as suggested by Figure~\ref{lumdis}a this leads to an overestimate of
the light at large radii, where the coverage is incomplete.

\section{Weak lensing signal}

Each galaxy gives only a noisy estimate of the weak lensing signal
because of its intrinsic shape. Therefore we average the shape
measurements of many sources to obtain a useful estimate of the
distortion $g$.  When we compute the ensemble averaged distortion, we
weight the contribution of each object with the inverse square of the
uncertainty in the measurement of the distortion as described in HFK.

\begin{figure}
\begin{center}
\leavevmode
\hbox{%
\epsfxsize=0.9\hsize
\epsffile{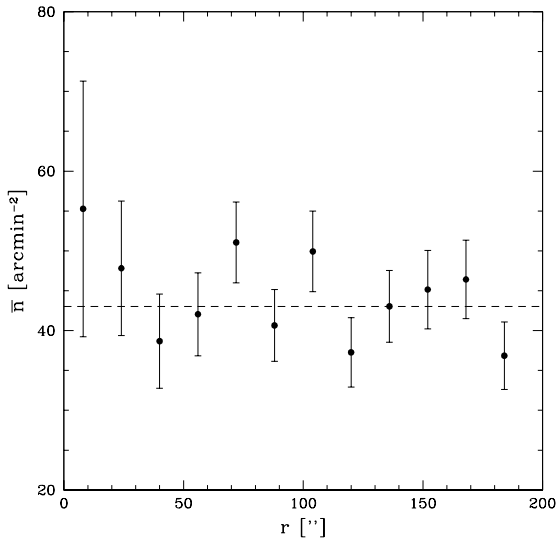}}
\begin{small}
\caption{Azimuthally averaged number density of background galaxies
($21.5<F814W<25.5$ and $0<F606W-F814W<1.4$) as a function of
distance from the cluster centre. The counts have been corrected
for the boundaries of the mosaic. No significant overdensity
is found at the cluster position, which indicates that the contamination
by cluster members is low. The dashed line shows the average number
density of 43 background galaxies arcmin.$^{-2}$
\label{profbg}}
\end{small}
\end{center}
\end{figure}

We select galaxies with $21.5<F814W<25.5$, and $0<F606W-F814W<1.4$ in
our sample of background galaxies. As mentioned above, we exclude
objects with sizes comparable to the PSF. The resulting sample
consists of 1130 galaxies, and has a median magnitude of $F814W=24.3$.
Comparison with Figure~\ref{colmag} shows that some faint cluster
members might end up in this sample of background sources. The
contamination will be most important in the central region of the
cluster. To estimate the contamination by cluster members we examined
the azimuthally averaged number density as a function of distance from
the cluster centre. The profile is presented in Figure~\ref{profbg}.
The number counts are slightly higher near the cluster centre, but 
the excess is not significant. In the further analysis we ignore
the data inside 40'' from the BCG.

The average number density of sources is 43 galaxies arcmin$^{-2}$.
Similar number densities can be reached in deep images taken from the
ground (e.g., B{\'e}zecourt et al. 2000). However, the main advantage
of our observations over ground based images is the much smaller
correction for the size of the PSF. As a result the lensing signal is
better calibrated, and the noise in the shape measurements from HST
data is lower.

\begin{figure*}
\begin{center}
\vspace{-0.5cm}
\leavevmode
\hbox{%
\epsfxsize=\hsize
\epsffile{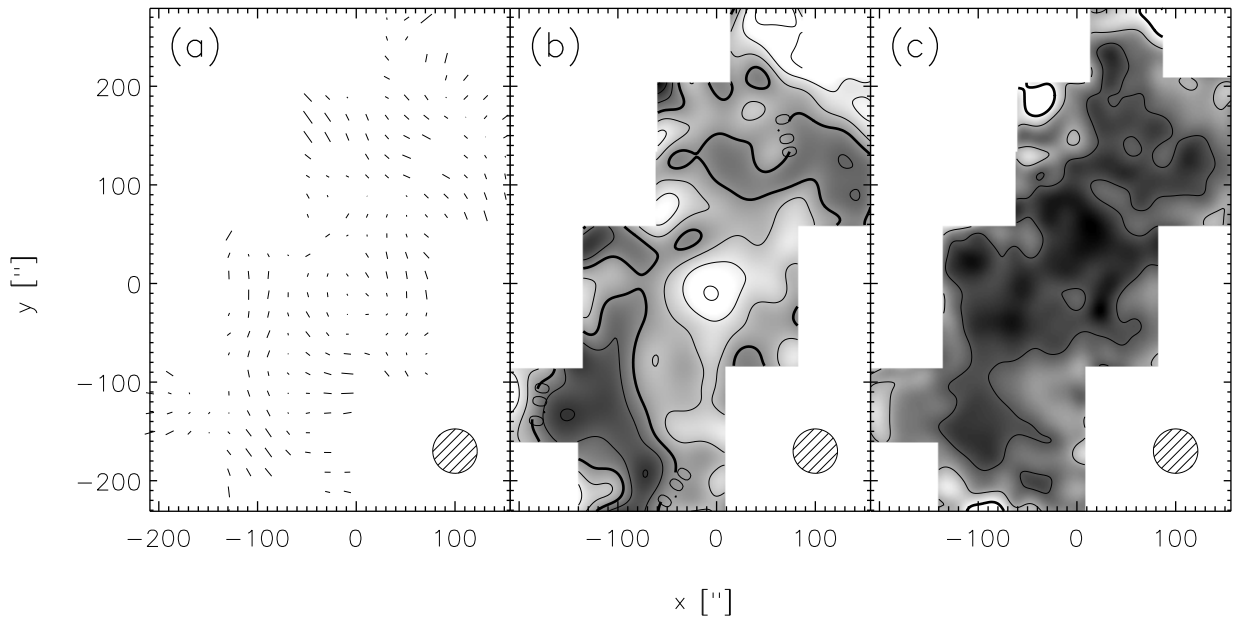}}
\begin{small}
\vspace{-0.3cm}
\caption{(a) Smoothed distortion field $g$ obtained using galaxies
with $22<F814W<26$ and $0<F606W-F814W<1.6$. The measurements have been
smoothed using a Gaussian with a FWHM of 45 arcseconds (indicated by
the shaded circle). The orientation of the sticks indicates the
direction of the distortion, and the length is proportional to the
amplitude of the signal. The origin of the plot coincides with the
assumed cluster centre. (b) The corresponding reconstruction of the
projected surface mass density. The interval between adjactent
contours is 0.05 in $\kappa$. (c) Noise map of the mass reconstruction
from bootstrapping resampling of the shape measurements. It shows that
the noise in the reconstruction increases rapidly towards the edges of 
the observed field. The interval between adjacent contours is 0.025.
The peak in the mass reconstruction is detected at the $3\sigma$ level.
\label{massrec}}
\vspace{-0.5cm}
\end{small}
\end{center}
\end{figure*}

Figure~\ref{massrec}a shows the smoothed distortion field from
the sample of source galaxies (we used a Gaussian with a FWHM of $45''$).
The position of the BCG corresponds to the origin of the plot. 
A systematic tangential alignment of the sources with respect to the 
cluster centre can be observed.

We use the distortion field presented in Figure~\ref{massrec}a
to reconstruct the projected surface mass density. The mass
reconstruction has been computed using the maximum likelihood
extension of the original KS algorithm (Kaiser \& Squires 1993;
Squires \& Kaiser 1996). This algorithm has the advantage over direct
inversion methods that it can be applied to fields with complicated
boundaries, such as our mosaic.

Figure~\ref{massrec}b shows a grey scale image of the reconstructed
surface mass density. The peak in the mass distribution coincides with
the position of the BCG. A bootstrapping resampling of the shape
measurements enables us to compute the noise map of the mass
reconstruction, which is presented in Figure~\ref{massrec}c.  The
noise in the mass reconstruction increases rapidly towards the edges
of the observed field. From the noise map we find that the peak in the
mass distribution is detected at the $3\sigma$ level.

\section{Mass and mass-to-light ratio}

The azimuthally averaged tangential distortion $\langle g_T\rangle$ as
a function  of radius from the  cluster centre is a  useful measure of
the lensing signal (e.g., Miralda-Escud{\'e} 1991; Tyson \& Fischer 1995).
The tangential distortion is defined as $g_T=-(g_1\cos 2\phi + g_2\sin
2\phi)$,  where $\phi$  is the  azimuthal  angle with  respect to  the
assumed cluster centre, for which we take the position of the BCG.

\begin{figure}
\begin{center}
\leavevmode
\hbox{%
\epsfxsize=\hsize
\epsffile{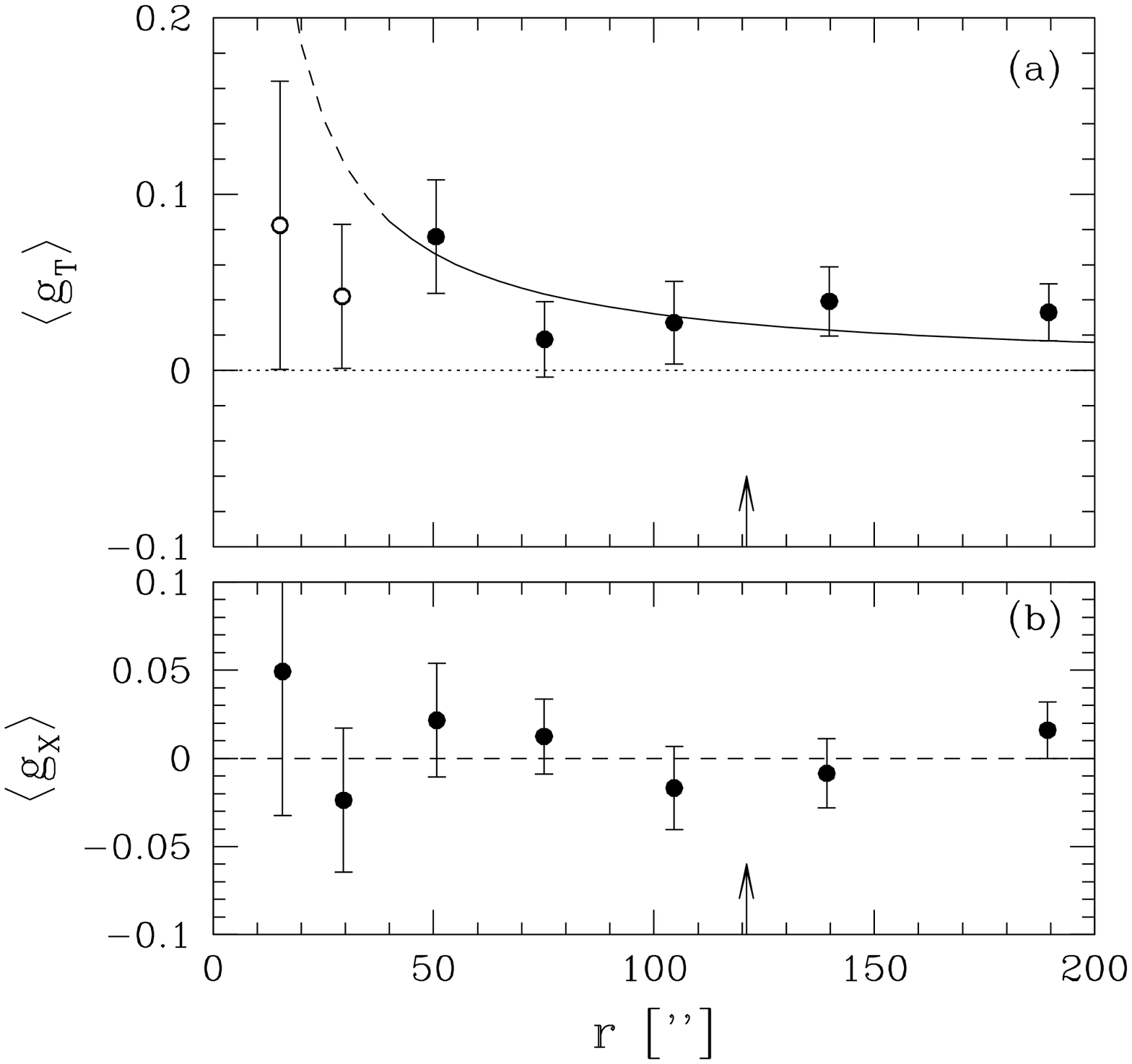}}
\begin{small}
\caption{(a) Average tangential distortion as a function of radius
from the cluster centre, for which we took the position of the
brightest cluster galaxy. The line corresponds to the profile of a
singular isothermal model ($\kappa(r)=r_E/2r$, where $r_E$ is the
Einstein radius) fitted to the data. Because of possible contamination
by faint cluster members we only fit to the data at radii larger than
40''. The best fit value for the Einstein radius is $r_E=6{\farcs}2\pm
1{\farcs}8$. (b) average signal when the phase of the distortion is
increased by $\pi/2$. If the signal shown in (a) is caused by
gravitational lensing, $\langle g_X\rangle$ should vanish, as is
observed. In both figures, the arrows indicate a radius of
$1~h_{50}^{-1}$ Mpc.\label{gtprof}}
\end{small}
\end{center}
\end{figure}

The azimuthally averaged tangential distortion as a function of radius
from the cluster centre is presented in Figure~\ref{gtprof}. A singular
isothermal sphere model ($\kappa(r)=r_E/2r$, where $r_E$ is the
Einstein radius) gives a best fitted $r_E=6{\farcs}3\pm1{\farcs}8$.
To minimize the diluting effect of cluster galaxies on the weak
lensing signal, we have excluded the measurements at radii smaller
than 40'' from the fit.

\subsection{Velocity dispersion}

The next step is to relate the measurement of the  Einstein radius
to a velocity dispersion, for which we use photometric redshift
distribution from the the northern and southern Hubble Deep
Fields (Fern{\'a}ndez-Soto, Lanzetta, \& Yahil 1999; Chen et al.
1998). HFK examined the usefulness of photometric redshift
distributions to calibrate the lensing signal and found that
they work well. 

The amplitude of the lensing signal as a function of source redshift
is characterized by $\beta$, which is defined as
$\beta=\max[0,D_{ls}/D_s]$, where $D_{ls}$ and $D_s$ are the angular
diameter distances between the lens and the source, and the observer
and the source. To compute $\langle\beta\rangle$ we also take into
account that fainter galaxies are noisier and have a lower weight in
the average. For our sample of sources we obtain
$\langle\beta\rangle=0.29$ (taking $\Omega_m=0.3$, and
$\Omega_\Lambda=0$).  Placing the background galaxies in a single
source plane at $z=1$ would yield a similar
$\langle\beta\rangle$. Using these results we derive a velocity
dispersion of $\sigma=886^{+121}_{-139}$ km/s.

The value of $\langle\beta\rangle$ does not only depend on the
redshifts of the sources, but it also depends on the cosmological
parameters that define the angular diameter distances. For an
$\Omega_m=1$, and $\Omega_\Lambda=0$ model we find essentially the
same $\langle\beta\rangle$, and $\sigma=881^{+120}_{-138}$ km/s. In a
$\Omega_\Lambda$ dominated universe the changes are larger. Assuming
$\Omega_m=0.3$, and $\Omega_\Lambda=0.7$ gives
$\langle\beta\rangle=0.33$, and results in $\sigma=831^{+113}_{-130}$
km/s.

The result from the weak lensing analysis is in excellent agreement
with the observed velocity dispersion of $817\pm80$ km/s, which
was determined from the velocities of 52 cluster members
(Van Dokkum et al. 2001, in preparation).

Clowe (1998) obtained deep $R$ band images of MS~2053 with the Keck
telescope, and measured the weak lensing signal. He used a source
redshift of $z=1.75$ and derived a velocity dispersion of 
$\sigma\sim 700$ km/s. Such high average source redshifts are
unrealistic, in particular when compared to the $z=1$ we use, based on
photometric redshift distributions. For more realistic redshift
distributions the result of Clowe (1998) increases to $\sigma\sim 900$
km/s, in good agreement with our results.

Luppino \& Gioia (1992) discovered a gravitationally lensed blue arc
in deep images of MS~2053. The arc is located approximately 15
arcseconds north of the BCG. To date, the redshift of the arc is not
known. If we assume a redshift of $z=2$ for the arc, and adopt a SIS
model (where the position of the arc gives the Eistein radius) the
corresponding velocity dispersion is about 1030 km/s. This value is
higher than, but consistent with the weak lensing estimate.  Moreover,
if the mass distribution is elongated in the direction of the arc the
strong lensing mass estimate is lowered.  Figure~\ref{lumdis}a
indicates that the light distribution is elongated, roughly in the
direction of the giant arc. Because of the low signal-to-noise ratio
of the weak lensing signal, we cannot constrain the elongation of the
mass distribution.

\subsection{Mass-to-light ratio}

From the sample of cluster galaxies we estimate a total cluster
luminosity of $(3.1\pm0.4)\times 10^{12}h_{50}^{-2}{\rm L}_{B\odot}$ within an
aperture of radius $1~h_{50}^{-1}$ Mpc (see section~4). The best fit
SIS model gives a projected mass of $(5.7\pm1.6)\times10^{14}
h_{50}^{-1}{\rm M}_\odot$ in the same aperture. Thus we obtain an
average mass-to-light ratio of $184\pm56~\moverl$ within
$1~h_{50}^{-1}$ Mpc.

Kelson et al. (1997) have studied the fundamental plane of MS~2053.
They find that the early type galaxies in the cluster define a 
clear fundamental plane. Comparison with low redshift clusters suggests 
that the structure of early type galaxies has changed little since
$z=0.58$. Similar analyses have been performed for other clusters
(e.g. van Dokkum \& Franx 1996; van Dokkum et al. 1998) and have
shown that the mass-to-light ratios of early type galaxies
evolve with redshift, which is accounted to luminosity evolution.

As a result also the global cluster mass-to-light ratios evolve with
redshift. The mass-to-light ratio of early type galaxies in MS~2053 in
the $B$ band is $37\pm4\%$ lower than present day values (Kelson et
al. 1997). Under the assumption that the total luminosity of the
cluster has changed by the same amount, we find an average
mass-to-light within $1~h_{50}^{-1}$ Mpc of $291\pm89\pm19~\moverl$,
corrected for luminosity evolution to $z=0$. The first contribution
to the error budget is the statistical uncertainty in the determination
of the mass-to-light ratio, and the second contribution is due
to the uncertainty in the correction for luminosity evolution.

\begin{figure}
\begin{center}
\leavevmode
\hbox{%
\epsfxsize=\hsize
\epsffile{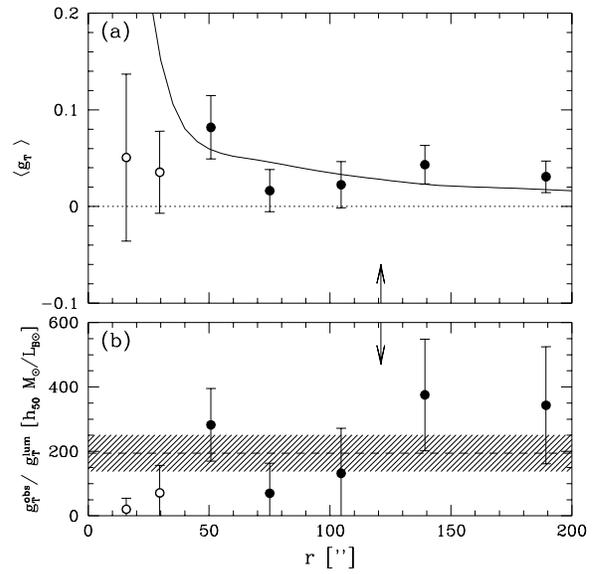}}
\begin{small}
\caption{(a) Plot of the average tangential distortion as a function
of radius from the BCG. The solid line is the expected tangential
distortion (scaled by the mass-to-light ratio to fit the observations)
derived from the average radial light profile, assuming the
mass-to-light ratio is constant with radius.  (b) The ratio of the
observed distortion and the derived distortion from the light (taking
$M/L_B=1$ in solar units). The shaded region indicates the one
$\sigma$ region around the average of the points with radii larger
than 40''.  The observations are consistent with a constant
mass-to-light ratio of $196\pm58~\moverl$. In both figures, the
arrows indicate a radius of $1~h_{50}^{-1}$ Mpc.
\label{moverl}}
\end{small}
\end{center}
\end{figure}

Under the assumption that the light traces the mass we can derive the
expected tangential distortion as a function of radius. To measure the
mass-to-light ratio, we scale the computed tangential distortion
$g_T^{\rm lum}$ to match the observed signal. In Figure~\ref{moverl}a
the resulting profile (solid line) is shown. The ratio of the computed
and observed signal is presented in Figure~\ref{moverl}b. Because of
possible contamination by faint cluster members we exclude the points
at radii less than 40'' from the fit. We find that the results are
consistent with a constant mass-to-light ratio with radius, and we
find an average value of $195\pm58~\moverl$.

\section{Combined results from rich clusters}

With the analysis of MS~2053 we have a sample of three clusters of
which the mass distribution has been studied using mosaics of WPFC2
images. For the analysis in section 7.3, we augment this sample with
the $z=0.22$ cluster Abell 2219, which has been studied from the
ground by B{\'e}zecourt et al. (2000).  Their analysis is identical to
ours, and like for our clusters, photometric redshifts have been used
to relate the lensing signal to the mass.

All clusters are X-ray selected, and their X-ray properties are listed
in Table~\ref{xraytab}.

\begin{table}
\begin{center}
\begin{tabular}{lcccc}
\hline
\hline
                & $z$  & $L_x$ (2-10 keV) 	& $kT$ 			& ref.	\\
		&      & [$h_{50}^{-2}~10^{44}$ ergs/s] & [keV]		&	\\
A~2219		& 0.22 & $38$			& $9.5\pm0.6$		& 1	\\
Cl~1358+62	& 0.33 & $11.4\pm0.3$	      	& $6.9\pm0.5$ 		& 2	\\
MS~2053-04	& 0.58 & $7.9\pm0.7$ 		& $8.1^{+3.7}_{-2.2}$   & 2	\\
MS~1054-03	& 0.83 & 28.6			& $12.4^{+3.1}_{-2.2}$  & 3	\\
\hline
\hline   
\end{tabular}
\begin{small}
\caption{X-ray properties of the three clusters for which we have
obtained HST mosaics, as well as the properties of A~2219. References:
(1) Allen (1998); (2) Henry (2000); (3) Donahue et al. (1998).
\label{xraytab}}
\end{small}
\end{center}
\end{table}

\subsection{Comparison between weak lensing mass and dynamical mass}

The large number of spectroscopic confirmed members in each of the
remaining three clusters results in accurate measurements of their
galaxy velocity dispersions. In this section we compare the weak
lensing estimates of the cluster velocity dispersions to the velocity
dispersion of the galaxies. In Table~\ref{comptab} we list the results
of the best fit SIS model to the observed weak lensing signal, as well
as the corresponding velocity dispersion inferred from lensing.

\begin{table*}
\begin{center}
\begin{tabular}{lcccccc}
\hline
\hline
                & $z$  & $r_E$ 		& $\langle\beta\rangle$ & $\sigma$ (WL) 	& $\sigma$ (galaxies)  & ref.\\
		&      & ['']		&			& [km/s]		& [km/s]	  & \\
Abell 2219	& 0.22 & $17.4\pm2.0$	& 0.46  & $1075^{+61}_{-63}$	& -	      & - \\
Cl~1358+62	& 0.33 & $10.8\pm1.4$   & 0.56	& $835^{+52}_{-56}$	& $910\pm54$  & 1 \\
MS~2053-04	& 0.58 & $6.3\pm1.8$ 	& 0.29	& $886^{+121}_{-139}$	& $817\pm80$  & 2 \\
MS~1054-03	& 0.83 & $11.5\pm1.4$	& 0.23	& $1311^{+83}_{-89}$	& $1150\pm90$ & 3 \\ 
\hline
\hline   
\end{tabular}
\begin{small}
\caption{Results from the weak lensing analyses of the three clusters
for which HST mosaics were obtained. References: (1) Carlberg et al. (1997);
(2) Van Dokkum et al. (2001, in preparation); (3) van Dokkum (1999).
\label{comptab}}
\end{small}
\end{center}
\end{table*}

For all three clusters we use the photometric redshift distribution
from the Hubble Deep Fields (Fern{\'a}ndez-Soto et al. 1999; Chen et
al.  1998). This lowered the $\langle\beta\rangle$ for Cl~1358
slightly compared to the value used in HFKS, and the new estimate for
the weak lensing velocity dispersion is listed in Table~\ref{comptab}.

The comparison with the velocity dispersions of cluster galaxies shows
a good agreement in all three cases, suggesting that the galaxy
velocity dispersions are characteristic of the cluster as a
whole. Similar comparisons have been made in the past (e.g., Smail et
al. 1997; Wu et al. 1998; Allen 1998).  The samples of clusters used
by Wu et al. (1998), and Allen (1998) are rather inhomogeneous:
different methods were used for the correction of the circularization
of the background sources, and no realistic redshift distributions of
the sources have been used. Wu et al. (1998) and Allen (1998) find in
general a fair agreement between the velocity dispersions of the
galaxies and the velocity dispersions derived from weak lensing
analyses.

\begin{figure}
\begin{center}
\leavevmode
\hbox{%
\epsfxsize=\hsize
\epsffile[10 150 550 600]{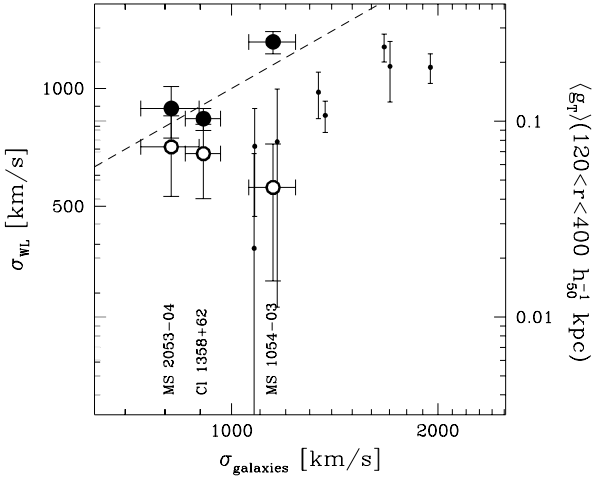}}
\begin{small}
\caption{Weak lensing velocity dispersion (as inferred from the best
fit SIS model) versus the observed velocity dispersion of the
galaxies. The large open points are the results from our HST mosaics,
measuring the distortion in an annulus $120<r<400~h_{50}^{-1}$
kpc. The results from Smail et al. (1997) are measured in the same
annulus and are indicated by the small dots. The large solid points
correspond to the measurements when the complete HST mosaics are
used. The line of equality between the weak lensing velocity
dispersion and the velocity dispersion of cluster galaxies is
indicated by the dashed line.  Our large field weak lensing results
agree well with this line. To allow a direct comparison with Figure~4
from Smail et al. (1997) we have indicated the average tangential
distortion in an annulus $120<r<400~h_{50}^{-1}$ on the right hand
axis.
\label{smail}}
\end{small}
\end{center}
\end{figure}  

A more systematic study was presented in Smail et al. (1997) who 
determined the weak lensing signal of 12 distant clusters observed
with WFPC2. Each cluster was observed with one pointing on the cluster
core. 

Smail et al. (1997) computed the average tangential distortion within
an annulus $120<r<400~h_{50}^{-1}$ kpc, and plotted the result against
the observed cluster dispersion. Under the assumption that the cluster
mass distribution is described by a SIS model, this measurement
provides an estimate for the velocity dispersion. Figure~\ref{smail}
shows the weak lensing estimate of the velocity dispersion versus the
velocity dispersion of cluster members for our sample. 

The large open points are the results from our HST mosaics, when we
confine the weak lensing analysis to the annulus used by Smail et
al. (1997). The large solid points show the results when the full HST
mosaic is used for the weak lensing analysis. The right hand vertical
axis displays the value of the average tangential distortion in the
annulus used by Smail et al. (1997), and allows a direct comparison
with their Figure~4. For comparison, we also show the results from
Smail et al. (1997) (small dots).

Smail et al. (1997) found a discrepancy between their weak lensing
signal and the velocity dispersion of the galaxies. They argued that
the velocity dispersions from the galaxies are overestimated by $\sim 40\%$,
compared to the velocity dispersion expected from the weak lensing analysis.
When we confine the weak lensing analysis of our HST mosaics to the
same annulus used by Smail et al. (1997) we too find that the velocity
dispersion inferred from the weak lensing analysis is lower than the
velocity dispersion of the galaxies. However, the results based on
the full mosaics agree well with the line of equality (dashed line).

Limiting the weak lensing analysis to the cluster core results in a
systematic underestimate of the cluster velocity dispersion. The
largest change is seen for MS~1054. For this cluster the explanation
is straightforward. HFK showed that the mass distribution in the
cluster centre is complex, consisting of three distinct clumps. As a
result the average tangential distortion is lowered, when only the
inner $400 h_{50}^{-1}$ kpc are considered in the analysis.

Several effects can introduce a systematic offset between the weak
lensing results and the dynamical measurements (e.g., Smail et al
1997). Because we find a good agreement between the two estimates
when the lensing signal is measured from wide field data, we
argue that it is the cluster mass profile in the core that gives
rise to the discrepancy. Only if the profile is isothermal, one
expects a good agreement, but substrucure or a shallower density
profile results lowers the lensing signal compared the value expected
from the SIS model.

\subsection{Average cluster mass profile}

Numerical simulations have indicated that dark matter halos
originating from dissipationless collapse of density fluctuations may
follow a universal density profile (e.g., Navarro, Frenk, \& White
1997). The Navarro, Frenk, \& White (NFW) profile appears to be an
excellent description of the radial mass distribution in these
simulations. The NFW profile is given by

\begin{equation}
\rho(r)=\frac{\delta_c\rho_c}{(r/r_s)(1+r/r_s)^2},
\end{equation}

\noindent where $\rho_c$ is the critical density of the universe,
$\delta_c$ is the characteristic overdensity, and $r_s$ is the scale
radius given by $r_s=r_{200}/c$, which all depend on the redshift and
mass of the halo. The parameter $c$ is referred to as the
concentration parameter. Given the cosmology, redshift, and mass of
the halo, $r_{200}$ follows immediately, and the values of $\delta_c$,
and $c$ can be computed using the routine {\tt CHARDEN} made available
by Julio Navarro\footnote{The routine {\tt CHARDEN} can be obtained
from {\tt http://pinot.phys.uvic.ca/${\tilde{\ }\!}$jfn/charden}}.

We have fitted the predicted profiles from NFW halos to the observed
tangential distortion of each cluster, and the best fit parameters are
listed in Table~\ref{tabnfw}. We have used a value of $\Gamma=0.18$
for the shape parameter of the CDM power spectrum. The parameter that
we fitted is $M_{200}$, the mass enclosed within a sphere of radius
$r_{200}$, and the other parameters are the ones produced by {\tt
CHARDEN} given $M_{200}$. The errors for $r_{200}$, $c$, and $r_s$
listed in Table~\ref{tabnfw} only reflect the uncertainty in these
parameters because of the uncertainty in the measurement of $M_{200}$.
We note that the resulting parameters are mainly determined by the
amplitude of the lensing signal (i.e. the mass of the halo) and not by
the shape of the density profile. Because of the strong substructure
in the centre of MS~1054 we excluded the measurements at radii less
than 75 arcsec.

\begin{table*}
\begin{center}
\begin{tabular}{lccccccccc}
\hline
\hline
(1)		& (2)  & (3)					 & (4)			& (5)	& (6)	 & (7)		      & (8)	 		     & (9)		  & (10) \\		
                & $z$  & $M_{200}$				 & $r_{200}$ 		& c	& $r_s$  & $\chi^2_{\rm NFW}$ & $P(\chi^2>\chi^2_{\rm NFW})$ & $\chi^2_{\rm SIS}$ & $P(\chi^2>\chi^2_{\rm SIS})$ \\
		&      & [$10^{14}h_{50}^{-1}$ M$_\odot$] & [$h_{50}^{-1}$ Mpc]	&	& [$h_{50}^{-1}$ kpc]	& & & & \\
A~2219		& 0.22 & $16.6^{+3.4}_{-3.2}$	 & $2.63^{+0.17}_{-0.18}$	& $5.01^{+0.13}_{-0.11}$	& $524^{+47}_{-48}$ & 8.6  & 0.74 & 10.0 & 0.62	\\
Cl~1358+62	& 0.33 & $2.8^{+0.8}_{-0.8}$	 & $1.36^{+0.12}_{-0.14}$	& $5.86^{+0.19}_{-0.14}$	& $231^{+27}_{-31}$ & 15.1 & 0.24 & 15.7 & 0.20	\\
MS~2053-04	& 0.58 & $2.8^{+1.6}_{-1.5}$	 & $1.18^{+0.19}_{-0.27}$	& $5.46^{+0.40}_{-0.22}$	& $217^{+46}_{-61}$ & 8.6  & 0.74 & 9.6  & 0.65	\\
MS~1054-03	& 0.83 & $14.4^{+2.8}_{-2.6}$	 & $1.81^{+0.11}_{-0.12}$	& $4.46^{+0.09}_{-0.08}$	& $407^{+32}_{-34}$ & 11.3 & 0.13 & 11.5 & 0.12	\\
\hline
\hline   
\end{tabular}
\begin{small}
\caption{Best fit parameters for the NFW profile (columns
3-6). The fitted paramter is $M_{200}$ and the other parameters are
computed using {\tt CHARDEN}, for an OCDM cosmology with
$\Omega_m=0.3$, $\Omega_\Lambda=0$, and shape parameter
$\Gamma=0.18$. The errors indicate the 68\% confidence intervals.
Columns~7 and 9 give the $\chi^2$ for the best fit NFW profile, and
SIS model, respectively. The probabilities of finding a larger value
for $\chi^2$ are listed in columns~8 and 10.
\label{tabnfw}}
\end{small}
\end{center}
\end{table*}   

\begin{figure}
\begin{center}
\leavevmode
\hbox{%
\epsfxsize=\hsize
\epsffile{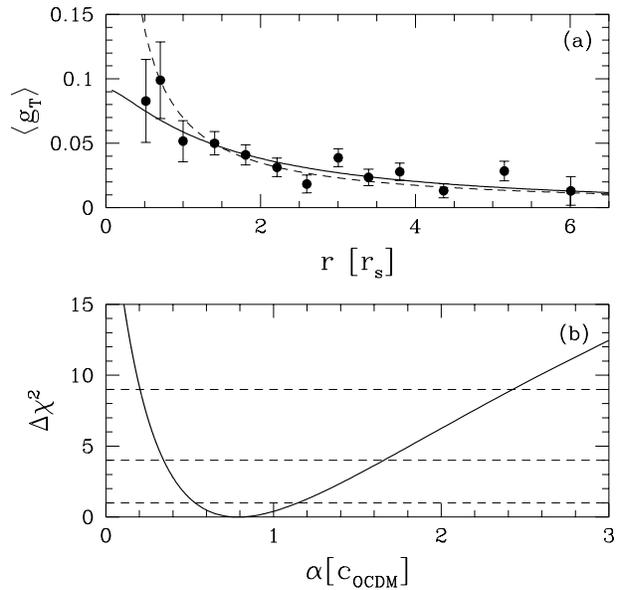}}
\begin{small}
\caption{(a) The ensemble averaged tangential distortion as a function
of radius (in units of $r_s=r_{200}/c$) for the four clusters in the
samples. We used the concentration parameters listed in
Table~\ref{tabnfw}.  The solid line is the best fit NFW profile, and
the dashed line is the best fit SIS model. The NFW provides the best
fit $(\chi^2=12.8)$, whereas the SIS model fit is worse
($\chi^2=17.1)$. (b) $\Delta\chi^2=\chi^2-\chi^2_{\rm min}$ as a
function of ratio between the measured and predicted concentration
parameter $\alpha/c_{\rm OCDM}$. The best fit NWF profile yields an
observed concentration parameter that is $0.79^{+0.44}_{-0.15}\times
c_{\rm OCDM}$
\label{nfw}}
\end{small}
\end{center}
\end{figure}

We now examine whether the NFW predictions match the actual
observations.  To do so, we scale the amplitude of the tangential
distortion profiles of the four clusters to the signal of a cluster
with $M_{200}$ of $5\times10^{14} h_{50}^{-1}$ M$_\odot$ at a redshift
$z=0.5$ (where we placed the sources at infinite redshift), and scale
the data radially in units of the derived value of $r_s$ (listed in
Table~\ref{tabnfw}), the scale length of the NFW profile. The
resulting ensemble averaged tangential distortion as a function of
$r_s$ is presented in Figure~\ref{nfw}a. This figure also shows the
best fit SIS model (dashed line) and the best fit NFW profile (solid
line).

The NFW profile provides a good fit to the data
$(\chi^2=12.8;~P(\chi^2>12.8)=0.38)$.  The SIS model fit is worse with
a $\chi^2=17.1$ $(P(\chi^2>17.1)=0.15)$.  The NFW model that is fitted
to the observations has a concentration parameter that is $\alpha\times
c_{\rm OCDM}$. Thus we test whether the predicted concentration
parameters agree with the observed lensing signal.  Figure~\ref{nfw}b
shows $\Delta\chi^2=\chi^2-\chi^2_{\rm min}$ as a function of $\alpha$ in
units of $c_{\rm OCDM}$. We find that the best fit value is
$0.79^{+0.44}_{-0.15}\times c_{\rm OCDM}$.  Thus the predicted
concentration parameter is in good agreement with the observations.

In the above, we have used NFW models for which the parameters were
obtained from the lensing data. Although the parameters are
essentially determined by the amplitude of the lensing signal, it is
useful to examine this in more detail. To do so, we use the observed
velocity dispersions of the galaxies to obtain an estimate of
$r_{200}$, using $r_{200}=(\sqrt{3}\sigma_{\rm gal})/(10 H(z))$ ,
where $H(z)$ is the value of the Hubble parameter at the redshift of
the cluster. We omit A~2219, because the galaxy velocity dispersion is
not known. We compute the NFW parameters and scale the tangential
distortion profiles of the three remaining clusters. The resulting
ensemble averaged profile is compared to the NFW profile in the same
way as before. We find that the NFW profile is a good fit
$(\chi^2=13.4)$, and that the SIS model fits worse
$(\chi^2=20.5)$. The best fit concentration parameter is found to be
$0.83^{+0.47}_{-0.37}\times c_{\rm OCDM}$. Thus both approaches yield
similar results.

These results shows that a systematic weak lensing study of a number
of clusters provides a direct way to test consistency of the predictions
of the theory of dissipationless collapse in CDM cosmologies.

\subsection{Cluster mass-to-light ratio}

Table~\ref{compml} lists the estimates of the average mass-to-light
ratio within an aperture of $1~h_{50}^{-1}$ Mpc radius. All values are
given in the restframe $B$ band.  Except for A~2219, the luminosity
evolution of the early type galaxies has been measured by studying the
fundamental plane (e.g., Kelson et al. 1997; Van Dokkum et al. 1998.)
Under the assumption that the global cluster mass-to-light ratio
evolves similarly with redshift we can correct the observed
mass-to-light ratio for luminosity evolution to $z=0$, and the results
are also listed in Table~\ref{compml}. The uncertainty in the
luminosity evolution results in an additional contribution to the
total error budget. In Table~\ref{compml} we list the statistical
error in the measurement of the mass-to-light ratio and the
uncertainty due to the correction for luminosity evolution separately.

\begin{figure}
\begin{center}
\leavevmode
\hbox{%
\epsfxsize=\hsize
\epsffile{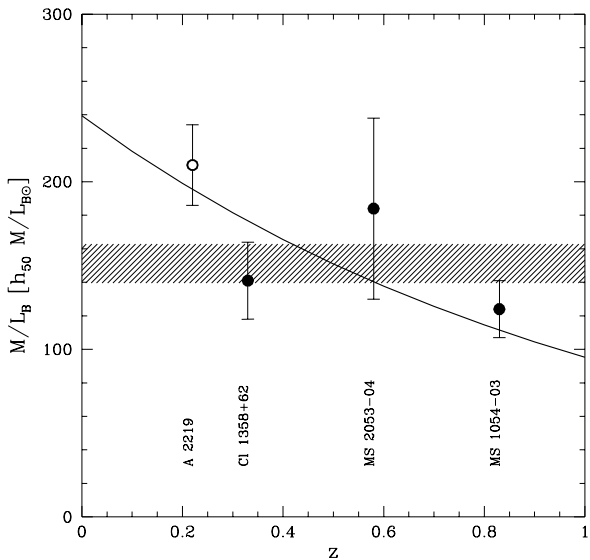}}
\begin{small}
\caption{The observed average mass-to-light ratio within a
$1~h_{50}^{-1}$ Mpc radius aperture of the clusters in the sample as a
function of redshift. The shaded region indicates the $1\sigma$ region
around the average mass-to-light ratio (assuming no luminosity
evolution). The assumption of an unevolving mass-to-light ratio is excluded 
at the 99\% confidence level. The solid line corresponds to the luminosity 
evolution as a function of redshift as inferred from studies of the 
fundamental plane of distant clusters of galaxies (Van Dokkum et al. 1998), 
and is consistent with the observations.
\label{evol}}
\end{small}
\end{center}
\end{figure}

Figure~\ref{evol} shows the observed average mass-to-light ratio in
the $B$ band within an aperture of $1~h_{50}^{-1}$ Mpc as a function
of cluster redshift. The dashed region in Figure~\ref{evol}
corresponds to the average of the observed mass-to-light ratios
(i.e. assuming no luminosity evolution), which yields a value of
$M/L_B=151\pm12\moverl$. The observations are inconsistent with an
unevolving cluster mass-to-light ratio that does not evolve at the 99\%
level.

\begin{table}
\begin{center}
\begin{tabular}{lccc}
\hline
\hline
                & $z$  	& ${\rm M}/{\rm L}_B$ (obs) 	& ${\rm M}/{\rm L}_B$ $(z=0)$ \\
		& 	& [\moverl]			& [\moverl]  \\
A~2219		& 0.22	& $210\pm24$			& $256\pm29\pm6$ \\
Cl~1358+62	& 0.33  & $141\pm23$			& $186\pm30\pm10$ \\
MS~2053-04	& 0.58  & $184\pm56$			& $291\pm89\pm19$ \\
MS~1054-03	& 0.83  & $124\pm17$			& $269\pm37\pm31$ \\
\hline
\hline   
\end{tabular}
\begin{small}
\caption{Average mass-to-light ratio within apertures of $1~h_{50}^{-1}$ Mpc
radius. The error budget of the mass-to-light ratios corrected for
luminosity evolution consists of the statistical uncertainty in the
measurement of the mass-to-light ratio (first error) and the contritbution
due to the uncertainty in the luminosity evolution (second error).
\label{compml}}
\end{small}
\end{center}
\end{table}

The solid line corresponds to the luminosity evolution as a function
of redshift as inferred from studies of the fundamental plane of
distant clusters of galaxies (Van Dokkum et al. 1998), scaled to fit
the observed total cluster mass-to-light ratios.  The evolution of the
cluster mass-to-light ratio of X-ray selected clusters is consistent
with the evolution of the mass-to-light ratio of the early type
galaxies. Van Dokkum et al. (1998) found that the $M/L_B$ ratio
evolves as $\Delta\log M/L_B\propto (-0.40\pm0.04)z$, which results in
an average value of $M/L_B=239\pm18\pm9\moverl$ for clusters at $z=0$.
The first error indicates the statistical uncertainty in the
measurement of the mass-to-light ratio, and the second error indicates
the additional error introduced by the uncertainty in the luminosity
evolution.

Carlberg et al. (1997) analysed a sample of 16 rich clusters, and also
found that the cluster mass-to-light ratios are consistent with a
universal value. They found an average value of $M/L_r=119 \pm
21~h_{50}{\rm M}_\odot/{\rm L}_{r\odot}$.  To convert this value to a
mass-to-light ratio in the $B$ band, we assume an average colour of
the cluster of $B-r=1.07$, which corresponds to the typical colour of
S0 galaxies (J{\o}rgensen et al. 1995).  Thus we find that the
estimate for the average cluster mass-to-light ratio from Carlberg et
al. (1997) corresponds to $219\pm38~\moverl$ (where we also corrected
for luminosity evolution to $z=0$), in excellent agreement with our
results.

Given the small spread in cluster mass-to-light ratios the
star formation efficiency in rich clusters appears to be a well regulated
process, although the sample of clusters needs to be increased before
firmer conclusions can be drawn.

\section{Conclusions}

We have presented the results of our weak lensing analysis of
MS~2053-04, a cluster of galaxies at a redshift $z=0.58$, for which we
detect a clear lensing signal. It is the third cluster we have studied
using a two-colour mosaic of deep WFPC2 images. Previously we have
studied Cl~1358+62 ($z=0.33$; HFKS) and MS~1054-03 ($z=0.83$; HFK).

The selected sample of background sources (with $21.5<F814W<25.5$ and
$0<F606W-F814W<1.4$) has a number density of 43 galaxies
arcmin$^{-2}$. Similar number densities can be reached in deep ground
based observations, but the correction for the circularization by the
PSF in WFPC2 images is much smaller. As a result the lensing signal
can be measured more accurately from space based images.

The position of the peak in the reconstruction of the cluster mass
surface density agrees well with the peak in the light distribution.
To measure the mass of the cluster we fit a SIS model to the observed
azimuthally averaged tangential distortion. The corresponding value
for the Einstein radius is $r_E=6{\farcs}2\pm 1{\farcs}8$. To relate
the Einstein radius to an estimate of the cluster velocity dispersion,
we use published photometric redshift distributions inferred from the
northern and southern Hubble Deep Fields. The best fit SIS model
corresponds to a velocity dispersion of $\sigma=886^{+121}_{-139}$
km/s, which is in excellent agreement with the observed velocity
dispersion of cluster galaxies of $817\pm80$ km/s.

We have analysed the weak lensing signal of 3 clusters using wide
field HST data, and we find that the velocity dispersion derived from
weak lensing agrees well with the velocity dispersion of the cluster
galaxies. This result differs from Smail et al. (1997) who compared
the weak lensing signal to the galaxy velocity dispersion using HST
observations of cluster cores. Based on our results we argue that the
discrepancy is caused by deviations from the SIS model in the inner
regions of clusters (substructure or a flatter profile).  To obtain an
accurate estimate of the weak lensing velocity dispersion wide field
data are necessary.

We use a sample of 4 clusters that have been analysed uniformly to
study the average cluster profile. The NFW profile fits the ensemble
averaged lensing signal well, and the predicted concentration
parameter is in good agreement with the observations: the observed
value is found to be $0.79^{+0.44}_{-0.15}$ times the predicted value
for an OCDM model.

The observed average mass-to-light ratio of MS~2053 within a
$1~h_{50}^{-1}$ Mpc radius aperture is $184\pm56~\moverl$.  We have
examined the mass-to-light ratios of the clusters in our sample, and
find that the results are inconsistent with a non-evolving universal
mass-to-light ratio. The measurements are consistent with a universal
mass-to-light ratio for rich, X-ray selected, clusters of galaxies
which evolves with redshift similarly to the luminosity evolution of
the cluster galaxies (e.g., Kelson et al. 1997; van Dokkum et
al. 1998). The average cluster mass-to-light ratio, corrected to
$z=0$, is found to be $M/L_B=239\pm18\pm9~h_{50}~\moverl$ (where the
first error indicates the statistical uncertainty in the measurement
of the mass-to-light ratio, and the second error is due to the
uncertainty in luminosity evolution), in good agreement with the
results from Carlberg et al. (1997) based on a dynamical study of 16
rich clusters.  The small spread in cluster mass-to-light ratios
suggests that the total star formation in clusters is a well regulated
process.

\section*{Acknowledgments}
P.G. v. D. was supported by Hubble Fellowship grant HF-01126.01-99A.
We would like to thank the Kapteyn Astronomical Institute for
their generous support.

\end{document}